\documentclass{article}

\usepackage{arxiv}

\usepackage[utf8]{inputenc} 
\usepackage[T1]{fontenc}    
\usepackage[colorlinks=true,
            linkcolor=blue,
            citecolor=blue,
            urlcolor=blue]{hyperref}
\usepackage{url}            
\usepackage{booktabs}       
\usepackage{amsfonts}       
\usepackage{nicefrac}       
\usepackage{microtype}      

\usepackage{graphicx}
\usepackage{doi}
\usepackage{amssymb}
\usepackage{amsmath}
\usepackage[version=3]{mhchem} 
\usepackage[T1]{fontenc}
\usepackage[utf8]{inputenc}
\usepackage{lmodern}      
\usepackage[american]{babel}
\usepackage{csquotes}
\usepackage{amssymb}
\usepackage{orcidlink}
\usepackage{verbatim}
\usepackage{graphicx}
\usepackage{float}
\usepackage[justification=centerlast]{caption}
\usepackage{subcaption}
\usepackage{tabularx}
\usepackage{booktabs}
\hypersetup{pdfauthor=author}
\usepackage{placeins}
\usepackage[multiple]{footmisc}
\usepackage[separate-uncertainty = true]{siunitx}
\usepackage[nolist]{acronym}
\usepackage{verbatim}
\usepackage{breakcites}
\usepackage[
  backend=biber,
  style=numeric,
  sorting=none,
  maxnames=2,
  minnames=1,
  sortcites=true,
  citestyle=numeric-comp,
  giveninits=true
]{biblatex}
\usepackage{authblk}
\AtEveryBibitem{%
  \ifentrytype{article}
    {\clearfield{issn}}
    {}%
}

\begin{acronym}
    \acro{AFM}{atomic force microscopy}
    \acro{TMM}{transfer matrix method}
    \acro{SSU}{spectro-spatial unmixing}
    \acro{SE}{spectroscopic ellipsometry}
    \acro{LE}{laser ellipsometry}
    \acro{NA}{numerical aperture}
    \acro{BF33}{BOROFLOAT 33 glass}
    \acro{AOI}{angle of incidence}
    \acro{PBR}{partial backside reflectance}
    \acro{PBRE}{partial backside reflectance - ellipsometry}
    \acro{PBRM}{partial backside reflectance - micro(spectro)scopy}
    \acro{PECVD}{plasma-enhanced chemical vapor deposition}
    \acro{FIB}{focused ion beam}
    \acro{HAADF}{high-angle annular dark-field}
    \acro{TEM}{transmission electron microscopy}
    \acro{STEM}{scanning transmission electron microscopy}
    \acro{EDX}{energy dispersive X-ray spectroscopy}
    \acro{MSE}{mean squared error}
    \acro{FWHM}{full width at half maximum}
    \acro{HOPG}{highly oriented pyrolytic graphite}
    \acrodefplural{AOI}{angles of incidence}
    \acro{SErr}{squared error}
    \acro{DFT}{density functional theory}
\end{acronym}

\title{Complex Refractive Index Determination via Microspectroscopy Through Magnifying Optics: Challenges and Opportunities}

\author[1]{Julian Schwarz\orcidlink{0000-0001-5194-6531}\thanks{Corresponding authors: \url{julian.schwarz@fau.de}, \url{mathias.rommel@iisb.fraunhofer.de},\\ \url{a.hutzler@fz-juelich.de}}}
\author[1]{Johannes Bauer\orcidlink{0009-0002-7509-6054}}
\author[$\ast$2]{Mathias Rommel\orcidlink{0000-0002-1141-3228}}
\author[3]{Andreas Hutzler\orcidlink{0000-0001-5484-707X}}

\affil[1]{Electron Devices, Friedrich-Alexander-Universität Erlangen-Nürnberg, Cauerstraße~6, \protect\\91058 Erlangen, Germany}
\affil[2]{Fraunhofer Institute for Integrated Systems and Device Technology IISB, Schottkystraße~10, \protect\\ 91058 Erlangen, Germany}
\affil[3]{Forschungszentrum Jülich GmbH, Helmholtz Institute Erlangen-Nürnberg for Renewable Energy (IET-2), \protect\\ Cauerstraße~10, 91058 Erlangen, Germany}

\renewcommand{\undertitle}{}

\hypersetup{
pdftitle={Complex Refractive Index Determination via Microspectroscopy Through Magnifying Optics: Challenges and Opportunities},
pdfsubject={physics.app-ph},
pdfauthor={Julian Schwarz, Johannes Bauer, Mathias Rommel, Andreas Hutzler},
pdfkeywords={Microspectroscopy, High numerical aperture, Van der waals materials, Refractive index, Material properties, Transmittance, Reflectance},
}

\begin{document}
\maketitle

\begin{abstract}
For the design and optimization of optoelectronic devices, accurate knowledge of the complex refractive indices of the constituent materials is essential.
Herein, we present a fast and non-destructive approach for the extraction of the refractive indices from reflectance and transmittance spectra of samples with lateral dimensions down to the micrometer scale.
Microspectroscopy, based on the combination of a standard optical microscope and a spectrometer, enables the assessment of the optical response of multilayer stacks using high-magnification optics with correspondingly large numerical apertures.
Employing a numerical formalism explicitly accounting for the influence of the numerical aperture, allows for precise retrieval of the refractive index without resorting to dispersion models.
We demonstrate the applicability of the proposed method for large-area, homogeneous, optically incoherent samples such as transparent glasses and absorbing 4H-\ce{SiC}, for a \ce{Si_xN_y} thin film on glass substrate, and for mechanically exfoliated flakes of highly oriented pyrolytic graphite and \ce{MoO3}, as representatives of uniaxial and biaxial optical anisotropy.  
While the results prove excellent agreement with values reported in literature, the case of graphite highlights the limitation for probing the out-of-plane refractive indices due to reduced sensitivity. 
Finally, we discuss possible extensions towards retrieving the full anisotropic tensor of the refractive index, establishing the proposed approach as a methodologically sound alternative to spectroscopic ellipsometry.
\end{abstract}

\keywords{Microspectroscopy \and High numerical aperture \and Van der waals materials \and Refractive index \and Material properties \and Transmittance \and Reflectance}

\section{Introduction}
The precise knowledge of the material properties is vital for the design and optimization of devices build from next-generation materials like van der Waals crystals, exploiting their unique optical and electronic features like tunable bandgaps~\cite{Chaves.2020}, high carrier mobilities~\cite{Li.2019}, and inherent anisotropy~\cite{Ermolaev.2021,Guo.2024,Li.2022c} for polarization-sensitive devices~\cite{Wang.2023,Wu.2024}.
For tailoring the characteristics in specific applications employing van der Waals materials like field-effect transistors~\cite{Li.2019,Wang.2023}, photodetectors~\cite{Jin.2023,Dodda.2022} and waveplates~\cite{AbediniDereshgi.2023,Lee.2021b}, the (complex) refractive index is a crucial parameter for selecting appropriate materials as it shapes the light-matter interaction in (opto)electronic devices.

The complex refractive index, consisting of the real part, the refractive index $n$, and the imaginary part, the extinction coefficient $k$, is commonly determined by \ac{SE}~\cite{Ermolaev.2020,Ermolaev.2021,Song.2018,Munkhbat.2022}, enabling the simultaneous determination of both quantities as well as the thickness~\cite{Chu.2020}.
However, the application of classical \ac{SE} is limited to large samples with lateral dimensions of several hundred micrometers or more~\cite{Munkhbat.2022,Luria.2020}. 
This requirement is often not met for flakes of van der Waals materials produced by mechanical exfoliation, whose typical lateral dimensions range from a few to a few tens of micrometers.
Hence, more specialized setups with smaller probe sizes are required for the characterization of such flakes, which can be achieved by imaging \ac{SE}~\cite{Wurstbauer.2010,Funke.2016}, microspot \ac{SE}~\cite{Kravets.2010,Isic.2011}, and Fourier plane micro-ellipsometry~\cite{Kenaz.2023,Chen.2021}, among others.
But, these advanced methods entail setups with increasing complexity, prolonged measurement times, or complex data analysis and calibration.
Moreover, \ac{SE} typically relies on optical dispersion models, such as the Drude--Lorentz~\cite{Jellison.2007,Toksumakov.2026}, Tauc--Lorentz~\cite{Munkhbat.2022,Ermolaev.2021}, and Sellmeier~\cite{AndresPenares.2021} formalism, especially for the simultaneous determination of the complex refractive indices and thicknesses of the materials under investigation.

An established technique enabling the optical characterization of samples on the micrometer scale is microspectroscopy~\cite{Frisenda.2017,Bing.2018,Niu.2018,Hutzler.2017, Hutzler.2019}, which can be easily implemented by coupling a spectrometer with a glass fiber to a standard optical microscope~\cite{Schwarz.2023}.
In this way the reflectance and/or transmittance spectra are assesed with a spot size down to \qtyrange{1}{2}{\micro\meter}~\cite{Niu.2018,Schwarz.2025b}.
From such measurements, parameters like the thickness~\cite{Schwarz.2023,Ivanova.2024,Puebla.2020} or the complex refractive indices~\cite{Slavich.2024c,Zhao.2020,Hsu.2019,Lee.2019} can be deduced by comparison with comprehensive modeling. 
If linearly polarized light is applied, even the anisotropic optical properties~\cite{Slavich.2024c,Guo.2024,Ross.2020,Lee.2021b,AbediniDereshgi.2023} and the orientation of the crystal axes can be determined.
However, the modeling is often based on the assumption of normal incidence, which may lead to significant errors in the extracted parameters for objective lenses with high \acp{NA} and the accompanying large range of \acp{AOI}.
Additionally, the utilization of transmittance measurements is underrepresented in literature, whereby especially a lack of detailed description of the measurement process for such measurements is evident.

For this reason, we present an in-depth study on the purely numerical extraction of the (complex) refractive index from reflectance and/or transmittance spectra obtained by microspectroscopy under arbitrary \ac{NA}.
Therefore, the extensive modular modeling framework from our previous works~\cite{Hutzler.2020,Schwarz.2023,Schwarz.2025,Schwarz.2025b,Schwarz.2026} is employed and refined to account for transmitted light.
In the course of this work, our objective lens dependent modeling for the influence of the \ac{NA}~\cite{Schwarz.2023} is transferred from reflectance to transmittance measurements, as well as the developed approach accounting for the partial detection of multiply reflected light in thick transparent substrates with incoherent behavior~\cite{Schwarz.2025b}.
On top of that, by incorporating linearly polarized light in the measurements and implementing it into modeling~\cite{Schwarz.2026}, we demonstrate the determination of the refractive indices along the different in-plane crystal axes of anisotropic materials.
Lastly, the chances and limitations of microspectroscopy for the accurate extraction of the full anisotropic tensor of refractive indices are discussed.
We start with the measurement and modeling procedure for reflectance and, specificially, transmittance spectra.
Building on this, we present extracted complex refractive indices for several thick incoherent substrates, both non-absorbing and absorbing, as well as for a \ce{Si_xN_y} thin film on a glass substrate, and flakes of the uniaxial van der Waals material \ac{HOPG} and the biaxial representative \ce{MoO3}.

\section{Materials and methods}

\subsection{Sample preparation and reference characterization}
\label{subsec:Sample preparation and reference characterization}
In this work, we investigated several optically thick substrates. 
We assumed purely specular reflection and transmission, as all samples were double side polished. 
The samples were \qty{500}{\micro\meter} thick BF33 glass wafers from Schott AG~\cite{BF33}, \qty{100}{\micro\meter} AF 32eco thin wafers from Schott AG~\cite{AF32eco}, \qty{500}{\micro\meter} thick \ce{Al2O3} windows from S.A.F.I.R., and \qty{500}{\micro\meter} thick \ce{SrTiO3} substrates from CRYSTAL GmbH. 
A \ce{Si_xN_y} thin film was deposited on a \ac{BF33} wafer as a single-wafer process using \ac{PECVD} at a deposition temperature of \qty{130}{\degreeCelsius} with \ce{NH3} and \ce{SiH4} as precursors.
The ca. \qty{500}{\nano\meter} thick layer was applied in two consecutive \qty{250}{\nano\meter} deposition procedures to increase the homogeneity of the layer thickness.
The measurements took place as-deposited without any further heat treatment.
These samples were already employed in a previous study~\cite{Schwarz.2025b}, where extensive reference characterization of \ac{FIB} prepared lamellae was performed by \ac{HAADF}-\ac{STEM} and \ac{EDX} and the thickness of the \ce{Si_xN_y} film was determined to be \qty{536.4}{\nano\meter}. 

The thickness of an n-doped 4H-\ce{SiC} wafer with the typical \qty{4}{\degree} off-axis cut normal to the c-plane was determined to \qty{351.5}{\micro\meter} as the average of 5 thickness measurements at different positions on the wafer obtained by a Logitech CG10 contact measurement gauge.
Flakes of \ac{HOPG} and \ce{MoO3} were mechanically exfoliated in atmosphere and transferred onto \qty{500}{\micro\meter} thick \ac{BF33} glass substrates.
The thickness of the flakes was determined by \ac{AFM} measurements using a Bruker Dimension ICON in intermittent-contact mode.
\subsection{Microspectroscopic measurements and modeling}
\label{subsec:Microspectroscopic measurements and modeling}

\subsubsection{Measurement setup and modus operandi}
\label{subsubsec:Measurement setup and modus operandi}

\begin{figure}[h!tbp]
\includegraphics[width=\linewidth]{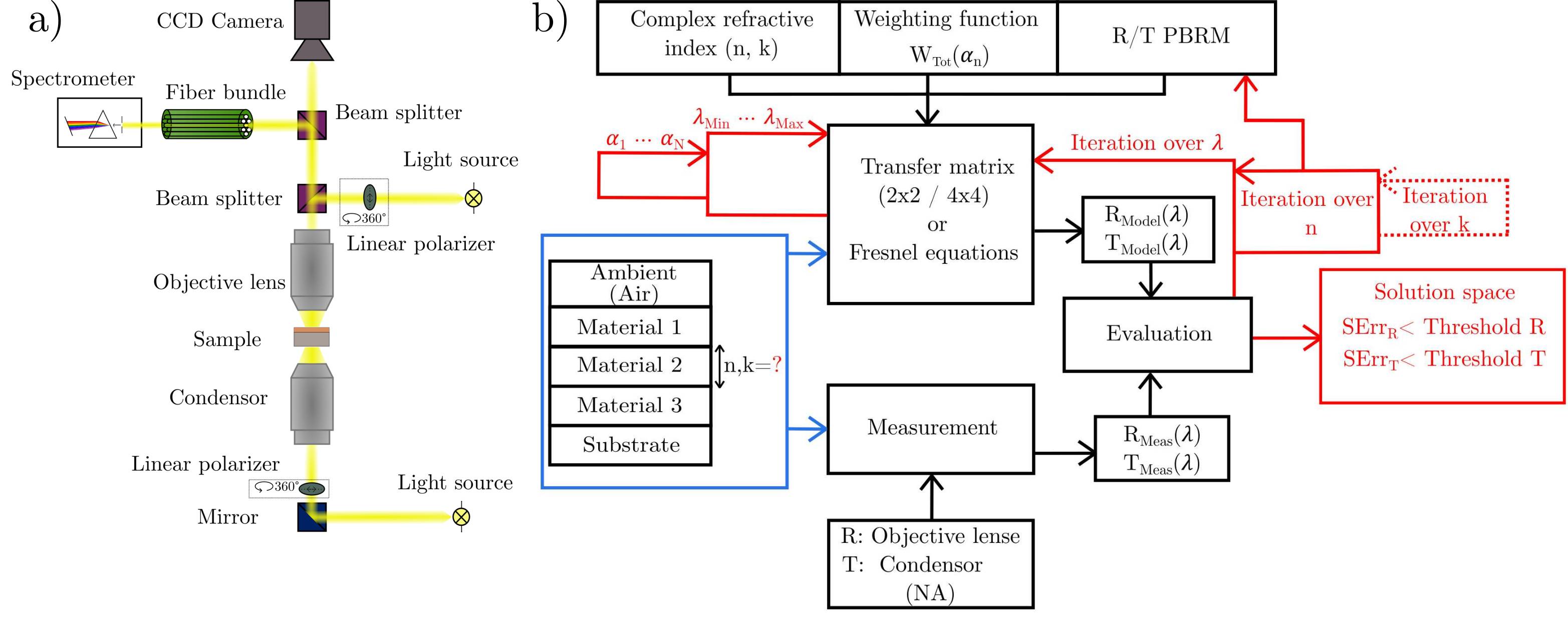}
\caption{a) Schematic of the employed measurement setup. b) Flowchart of the numerical procedure for extracting the (complex) refractive indices from measured reflectance and/or transmittance spectra.}
\label{Figure 1}
\end{figure}

The schematic of the employed measurement setup is shown in Fig.~\ref{Figure 1}a.
Reflectance and transmittance were measured with an Olympus BX53-MTRF microscope coupled to a Horiba iHR 320 spectrometer including a 1200 grooves/\qty{}{\milli\meter} diffraction grating.
The signal is collected with the center fiber of a fiber bundle, which has a diameter of \qty{200}{\micro\meter}.
Due to the characteristics of the halogen light sources and the Horiba Synapse CCD detector, the available spectral range is limited from \qty{440}{\nano\meter} to \qty{950}{\nano\meter}.
Linearly polarized light was achieved by the insertion of a continuously rotatable linear polarizer in the illumination path. 
For transillumination measurements, an achromatic and aplanatic Olympus U-AAC universal condenser was used. 
All utilized objective lenses are listed in Tab.~\ref{tab:Objective_lenses} together with their \acp{NA} and the corresponding diameters of the measurement spot on the sample, defined by the respective magnification and the fiber diameter.

\footnotetext[1]{'UL' stands for 'ultra large' working distance.}

\begin{table}[hbtp] 
\caption{Objective lenses, numerical apertures and diameters of the measurement spot.\label{tab:Objective_lenses}}
\newcolumntype{C}{>{\centering\arraybackslash}X}
\begin{tabularx}{\textwidth}{cCCCC}
\toprule
\textbf{Objective lens}	& \textbf{10x}	& \textbf{50xUL}\footnotemark[1]{}    & \textbf{100xUL}\footnotemark[1]{}    & \textbf{100x}\\
\midrule
\ac{NA}    & 0.25 & 0.35  & 0.60    & 0.90  \\
Spot diameter/\qty{}{\micro\meter} & 20 & 4 & 2 & 2\\
\bottomrule
\end{tabularx}
\end{table}

In general, the field stop was set to the smallest size to reduce stray light~\cite{Frisenda.2017,Schwarz.2025b}.
To account for background noise in reflectance, the spectrum of an ultra-low reflecting flock sheet from Musou Black was measured~\cite{Schwarz.2026}.
In transmittance, the background signal was isolated by measuring the spectrum with the light path to the detector blocked.  
For absolute intensity values, the reflectance spectra were referenced to those of a silicon sample.
In contrast, a \qty{100}{\percent} transmittance reference was used by aligning condenser and objective lens relative to each other without a sample in between.
All measurements were averaged over 10 acquisitions.

For focusing on the upper surface of the sample in both illumination modes, the objective lens was focused by moving the stage until the image of the field diaphragm was brought into sharp focus.
In transmission mode, the condenser was subsequently adjusted so that the image of its field diaphragm was also sharply visible, after the light source in the epi-illumination path had been disabled.
In general, the aperture diaphragm was fully opened in epi-illumination and in transillumination the condenser aperture was adjusted to match the \ac{NA} of the objective lens to ensure the same range of \acp{AOI} in both illumination modes.
However, the illumination through a thick substrate by the condenser causes spherical aberrations due to the refractive index mismatch between the substrate and the surrounding air~\cite{Everall.2010,OpticalSocietyofAmerica.1995b}.
This phenomenon is especially pronounced for the 100x objective lens with the highest \ac{NA} of 0.90 and the ocurring high \acp{AOI}, which is why the condenser aperture was reduced to an effective \ac{NA} of 0.60 for the transmittance measurements with this objective lens to mitigate the influence of spherical aberrations (cf. section 1 and Fig. S1 in the supporting information).

\subsubsection{Modeling framework}
\label{subsubsec:Modeling framework}

Fig.~\ref{Figure 1}b shows the flowchart of the numerical procedure for extracting the (complex) refractive indices from measured reflectance and  transmittance spectra.
In principle, the procedure is the inversion of the modeling process for the established method for the determination of the layer thickness from unpolarized reflectance spectra~\cite{Schwarz.2023} or the orientation of the crystal axes from linearly polarized reflectance spectra~\cite{Schwarz.2026}, when the refractive indices are known.
Here, in the reverse direction, the (complex) refractive indices are extracted from the measured spectra by comparison with the modeled spectra for a given set of the parameters refractive index $n$ and extinction coefficient $k$, which is done by iterating over the parameters and minimizing the \ac{SErr} between the measured and modeled intensity at the wavelengths of interest.
To reduce the computational effort for this process, the measurement data, originally recorded with high spectral resolution in the investigated wavelength range from \qtyrange{440}{950}{\nano\meter}, were interpolated to integer wavelengths with a step size of \qty{1}{\nano\meter}.
A requirement for the determination of the refractive indices is the knowledge of the sample structure, i.e., layer order, and the thickness of all layers as well as the refractive indices of the other materials besides the one of interest.
The thickness of other layers than the one of interest can also be determined via microspectroscopy, provided that there are isolated regions for each material available~\cite{Schwarz.2023}.

The influence of the \ac{NA} was considered by our objective lens dependent weighting function $W_{\mathrm{Tot}}$.
For this purpose, the distribution of \acp{AOI} is discretized in $N$ independent angles $\alpha_n$ and the individual spectra for these \acp{AOI} are weighted as per the definition of $W_{\mathrm{Tot}}$.
$W_{\mathrm{Tot}}$ comprises a geometric component based on the objective lens's \ac{NA} and a lens-dependent Gaussian shaped component, defined by the parameter $\zeta$, accounting for potential inhomogeneities of the intensity between the individual \acp{AOI}.
An elaborated derivation of the equations for $\alpha_n$ and $W_{\mathrm{Tot}}$ can be found in our previous work~\cite{Schwarz.2023}.
A discretization of $N=10$ was utilized since no further changes in the modeled spectra are apparent for a finer discretiziation and the thickness of the studied materials, as investigated previously~\cite{Schwarz.2025b}.
Eq.~\ref{eq:Reflectance} yields the resulting weighted reflectance~\cite{Schwarz.2023}:

\begin{equation}
    R(\lambda) = \sum_{n=1}^{N=10} R(\lambda,\alpha_{n}) \cdot W_{\textrm{Tot}}(\alpha_{n}).
\label{eq:Reflectance}
\end{equation}

If thick transparent substrates are used -- as required for transmittance measurements -- the displacement of multiply reflected light in the substrate relative to the directly reflected or transmitted light has to be considered.
Besides the incoherent behavior without interference fringes for thick substrates, multiply reflected light is either completely, partially, or not at all detected.
In order to account for this, we apply our developed approach, denoted \ac{PBRM}~\cite{Schwarz.2025b}, to both reflectance and transmittance.
The \ac{PBRM} technique is based on the \ac{AOI}, the thickness of the thick substrate and its refractive index as well as the geometric dimensions of the illuminated area on the sample and the measurement spot.
When the substrate is the material of interest for determining the refractive index, the respective iteration value for $n$ is fed in the calculation of the detected percentage of multiply reflected light.  
From our previous \ac{PBRM} study of the involved substrates at the present setup, we infer that for the 10x objective lens all orders of multiply reflected light are collected.
In contrast, for the objective lenses with higher magnifications only the directly reflected light (zeroth order) is detected~\cite{Schwarz.2025b}.
As the relations between the zeroth order and higher orders don't change for transmission measurements, we adopt the aforementioned information for transmittance modeling.
Solely the dimensions of the illuminated area, i.e., the image of the condenser's field stop, doesn't scale with the objective lens contrary to reflectance measurements.
However, we neglect this effect as a first order approximation. 

We employed 4×4 \ac{TMM} modeling in our preceding works~\cite{Schwarz.2023,Schwarz.2025b,Schwarz.2025,Schwarz.2026}, but without major changes, \ac{TMM}-based approaches are only capable of considering all orders of multiply reflected light~\cite{Schwarz.2025b}.
In reflectance, this shortcoming can be circumvented in an approximation by factoring in the substrate either as a layer or the substrate in the simulations~\cite{Schwarz.2025b}.
However, for modeling only the directly transmitted component the error introduced due to taking into account all orders of multiply reflected light might be significant.
While the induced error for the objective lenses other than the 10x lens in the case of \ac{BF33}, AF 32eco, and \ce{Al2O3} lies below \qty{0.5}{\percent}, the error for \ce{SrTiO3} amounts to ca. \qty{2}{\percent}.
As a consequence, Fresnel-based equations were used to simulate the individual $R(\alpha_n)$ and $T(\alpha_n)$ to accurately reproduce the detected multiply reflected light, since this approach allows the different orders of reflected and transmitted light to be separated in the calculations~\cite{Heavens.1991}.

In the case of thickness determination, the iteration over the thickness of the material of interest and minimization of the error between measurement and model yields the right solution, assuming an ideal constellation.
Due to ambiguity, in the reverse process however, multiple solutions for the (complex) refractive indices are possible.
Beyond purely numerical artifacts, interference effects in thin films introduce a thickness- and \ac{AOI}-dependent ambiguity.
Thus, even methods that are less complex from a modeling perspective and employ only a single \ac{AOI}, such as UV--Vis spectroscopy, suffer from multiple branches of solutions~\cite{Yin.2013}.
Since the simulated reflectance spectra under normal incidence of a hypothetical layer already exhibit intersections for varying refractive indices at a fixed extinction coefficient, and vice versa, ambiguities arise when extracting these parameters from such spectra (cf. section S2 and Fig.~S2 in the supporting information).
As a result, we implement threshold levels for reflectance and transmittance modeling, where all solutions below these \ac{SErr} values are allowed.

Furthermore, using only one measured physical quantity -- either reflectance or transmittance -- widens the solution space for $n$ and $k$ drastically.
While this is no obstacle for lossless media with $k=0$, extracting the complex refractive indices of media with non-negligible absorption requires a minimum of two independent measurements for each wavelength.
In literature, an approach is to utilize measurements of the same material with the same thickness on different substrate materials or substrate thicknesses, e.g., a monolayer of a van der Waals material on \ce{SiO2}/\ce{Si} substrates with varying oxide thicknesses~\cite{Zhang.2015,Hsu.2019}. 
Alternatively, the transmittance is another less commonly investigated measurement quantity next to the frequently used reflectance.
However, there are no studies on the exact modus operandi for measuring and modeling of the transmittance under arbitrary \ac{NA}.
Therefore, we transferred our modeling approach for \ac{NA} correction via $W_{\mathrm{Tot}}$ as per Eq.~\ref{eq:Reflectance} to the analysis of transmittance data and tested the capabilities at the example of thickness determination for the \ce{Si_xN_y} thin film on \ac{BF33} substrate with the objective lenses in Tab.~\ref{tab:Objective_lenses}.
All applied $\zeta$ parameters for the objective lenses can be found in Tab.~S1 in the supporting information. 
The results for the \ce{Si_xN_y} thin film are elaborated on within section 3 along Fig.~S4 and Tab.~S2 in the supporting information, showing similar accuracy to the well established extraction from reflectance spectra.

With the demonstration of precise transmittance modeling for arbitrary \acp{NA}, the flowchart in Fig.~\ref{Figure 1}b is concluded.
This enables the extraction of the (complex) refractive indices of conducting media from reflectance and transmittance spectra in microspectroscopy.
For non-absorbing materials the isolated use of reflectance or transmittance measurements is sufficient.
It should be noted here, that for uniaxially anisotropic layers, only an isotropic approximation can be determined.
This is because, in this case, four independent measured physical quantities need to be available for the four unknown parameters (complex refractive index parallel to the optical axis and perpendicular to it).
For biaxial materials, the use of linearly polarized light along the crystal axes in the xy-plane allows, at least, a uniaxial configuration to be achieved, thereby enabling an isotropic approximation along both axes in the xy-plane.
Based on our previous work, we conclude that, due to the rotationally symmetric illumination occuring in microspectroscopy, the linearly polarized spectra along a crystal axis in the xy-plane can be treated like unpolarized light with the refractive indices along the specificied crystal axis.
This contrasts with the expectation that, in the case of linearly polarized light, the large \acp{AOI} that occur at high \ac{NA} result in purely p- or s-polarized light.~\cite{Schwarz.2026}.

By using threshold values for the \ac{SErr}, solution spaces for $n$ and $k$ are extracted rather than single-valued estimates. 
The mean values of $n$ at each wavelength were then smoothed using Savitzky--Golay filtering~\cite{Savitzky.1964}. 
These curves are hereafter referred to as the smoothed mean. 
In line with the previously described interpolation to a wavelength grid with \qty{1}{\nano\meter} spacing, the window length was set to 51, corresponding to a spectral window of \qty{51}{\nano\meter}, and the polynomial was of the third order. 
For materials that are transparent in the investigated wavelength range, a first-order Sellmeier model according to Eq.~\ref{eq:Sellmeier} was fitted to the solution spaces to account for the dispersion of $n$~\cite{Sellmeier.1872}. 
To quantify the uncertainty, 100,000 iterations were performed, each using random values of $n$ from the solution space for each wavelength, and from this the mean values for B and C were extracted along with their standard uncertainties.

\begin{equation}
   n^2 = 1+\frac{B\lambda^2}{\lambda^2-C}
\label{eq:Sellmeier}
\end{equation}

For reference modeling of the reflectance and transmittance spectra, suitable (complex) refractive indices for air~\cite{Peck.1972}, Si~\cite{Green.2008}, \ce{Si_xN_y}~\cite{Vogt.2015}, \ce{Al2O3}~\cite{Weber.1986}, AF 32eco glass~\cite{AF32eco}, \ac{BF33}~\cite{BF33}, and \ce{SrTiO3}~\cite{Weber.1986} were taken from literature.
Despite the inherent uniaxial anisotropy, we modeled the crystalline \ce{Al2O3} sample as isotropic with material properties according to those of the c-plane.
This is justified based on the negligible birefringence with $\Delta n\approx0.08$ at \qty{500}{\nano\meter}~\cite{Weber.1986,Schwarz.2023}.
Similary, for the inherently uniaxial 4H-\ce{SiC} sample, an isotropic approximation for the refractive indices was applied~\cite{Khadivianazar.2019}.
This simplification is justified as an approximation due to the 10x objective lens used for the measurement, which has a low \ac{NA} of 0.25, and the correspondingly small \acp{AOI} inducing a diminishing component of the electric field oscillating in the out-of-plane direction~\cite{Schwarz.2023}.
For flakes composed of the van der Waals materials \ac{HOPG} and \ce{MoO3}, anisotropic refractive indices were chosen from literature~\cite{Boosalis.2015,AbediniDereshgi.2023}, which precisely described the material properties of flakes exfoliated by us in previous studies~\cite{Schwarz.2023,Schwarz.2025,Schwarz.2026}.

\section{Results and discussion}

\subsection{Incoherent samples}

Thick transparent samples exhibiting incoherent behavior are the least complex material class in this study.
Section 4 of the supporting information contains the results for 4 exemplary representatives of such samples.
The findings for \ac{BF33} in Fig.~S5, AF 32 eco in Fig.~S6, \ce{Al2O3} in Fig.~S7, and \ce{SrTiO3} in Fig.~S8 demonstrate the feasibility of the proposed method.
So, precise refractive indices matching the reference dispersions were extracted for these nonabsorbing materials for the 10x and 100x objective lenses.
In general, $n$ was usually determined solely from the reflectance spectra, since for $k=0$, a single measured quantity is sufficient to determine $n$.
For \ac{BF33}, Fig.~S5 also shows the determination of $n$ from the combination of reflectance and transmittance.
However, due to sporadic shifts in the measured transmittance towards lower values -- likely caused by uncertainty in the mechanical focusing of the condenser -- this method may result in slightly larger errors than when using reflectance data alone.

\begin{figure}[H]
\includegraphics[width=\linewidth]{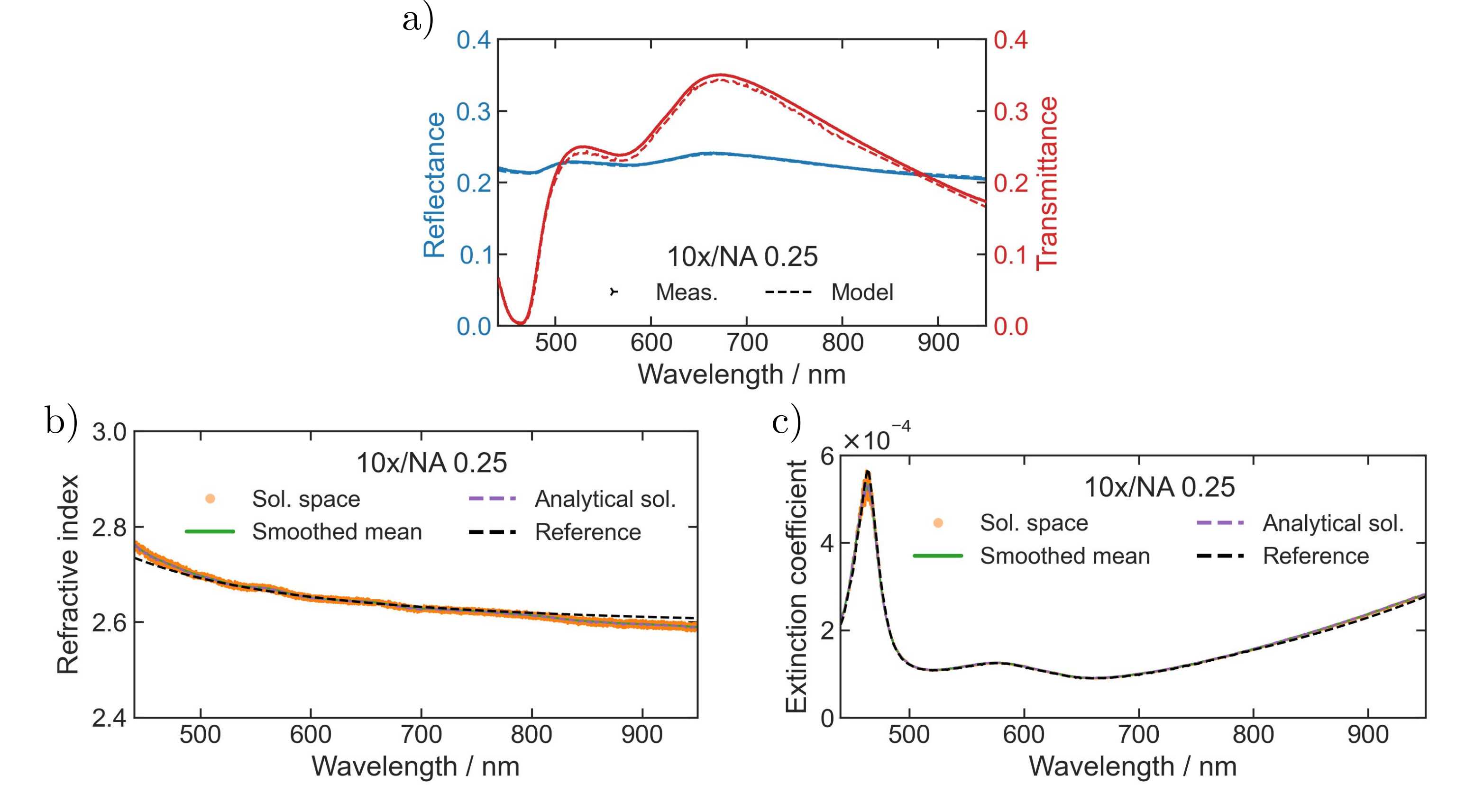}
\caption{(a) Measured reflectance ($R$) and transmittance ($T$) spectra of a \qty{351.5}{\micro\meter} thick 4H-\ce{SiC} sample for the 10x/\ac{NA}~0.25 objective lens and corresponding modeled spectra, calculated using isotropic optical constants ($n$, $k$) from~\textcite{Khadivianazar.2019} as reference. Extracted solution space, smoothed mean of the solution space, analytical solution based on the equations of~\textcite{EnricoNichelatti.2002}, and the reference values by~\textcite{Khadivianazar.2019} for the (b) refractive indices $n$ and (c) extinction coefficients $k$.}
\label{Figure 2}
\end{figure}

On the contrary, for thick absorbing samples, reflectance and transmittance are required to unambiguously determine $n$ and $k$. 
As an example, a c-plane oriented 4H-\ce{SiC} sample was investigated, whereby the \qty{4}{\degree} off-axis cut was neglected.
Fig.~\ref{Figure 2}a shows the measured and modeled reflectance and transmittance spectra obtained with the 10x/\ac{NA} 0.25 objective lens.
For the simulations, the reference thickness of \qty{351.5}{\micro\meter} was used, together with complex refractive indices from literature derived from similar samples~\cite{Khadivianazar.2019}.
The simulations accurately replicate the measurements.

\FloatBarrier

Figs.~\ref{Figure 2}b and c show the solution spaces for $n$ and $k$ extracted from the measurement data in Fig.~\ref{Figure 2}a.
The \ac{SErr} thresholds were set to \qty{1e-6}{} and the iteration ranges for $n$ and $k$ were defined from 2 to 3 and \qtyrange{0}{0.001}{} with steps of 0.001 and \qty{1e-6}{}, respectively. 
The Savitzy-Golay-filtered mean values lie very close to the reference dispersion.
Especially, the \ce{N}-doping associated absorption peak around \qty{460}{\nano\meter} is accurately reproduced in Fig.~\ref{Figure 2}c~\cite{Firsov.2016}.
Furthermore, the results of an analytical solution based on equations assuming normal incidence coincide perfectly with the smoothed mean values~\cite{EnricoNichelatti.2002}.
In this case, the analytical solution is applicable since the low \ac{NA} of 0.25 can be reasonably approximated by normal incidence, in particular for incoherent samples without \ac{AOI}-dependent interference patterns. 

\subsection{Thin films and van der Waals materials}

After the successful demonstration of the method for thick incoherent samples, homogeneous thin films exhibiting interference patterns are the next representative with increased complexity.
By the example of the \ce{Si_xN_y} thin film on a \ac{BF33} substrate, already utilized for thickness determination from transmittance spectra, the capabilities of the proposed approach are proven in section S5 in the supporting information.
There, Fig.~S9 presents the precise extraction of the refractive indices by combining the solutions obtained from the reflectance and transmittance spectra for the 10x and 100x objective lenses.
By analyzing the overlap of the solution spaces of both objective lenses, the ambiguity of multiple NA-dependent side branches is resolved, as these features are identified as numerical artifacts arising from the NA-dependent interference pattern, superimposed with measurement and modeling uncertainties.
The adaption of a Sellmeier model yields excellent agreement with the reference dispersion from literature~\cite{Vogt.2015}. 
Again, the infrequently arising shift in the measured transmittance towards lower values causes uncertainties in the determined behavior of the extinction coefficient.
Hence, a slightly increased $k$ is obtained.
However, the extinction coefficient is typically of the order of \qty{e-2}{} for \ce{Si_xN_y} thin films~\cite{Vogt.2015}, thus, its contribution is comparatively small for typical thin films.
Furthermore, a comparison based on additional refractive index profiles for \ac{PECVD}-deposited \ce{Si_xN_y} films reported in literature shows that the measured dispersion fits very well within this range~\cite{BakerFinch.2011,Duttagupta.2012,Wan.2013,Vogt.2015,Beliaev.2022}.
\FloatBarrier

The exploitation of the \ac{NA}-dependent behavior is only feasible for homogeneous layers covering a large area.
However, for mechanically exfoliated flakes of van der Waals materials the measurement spot of objective lenses with low magnification is not sufficient for covering only the material of interest.
Although our previously established \ac{SSU} approach is capable of distinguishing signals from multiple layer stacks within the measured area~\cite{Schwarz.2025}, the exclusive use of high-magnification objective lenses, where only the material of interest exists within the measured area, is advantageous for meeting the stringent accuracy requirements for refractive index determination.
Hence, only the 100x objective lens was utilized for the following flakes.
Based on the results in section~\ref{subsubsec:Measurement setup and modus operandi} and section S1 of the supporting information, the effective \ac{NA} was set to 0.90 for reflectance and 0.60 for transmittance measurements.

\begin{figure}[H]
\includegraphics[width=\linewidth]{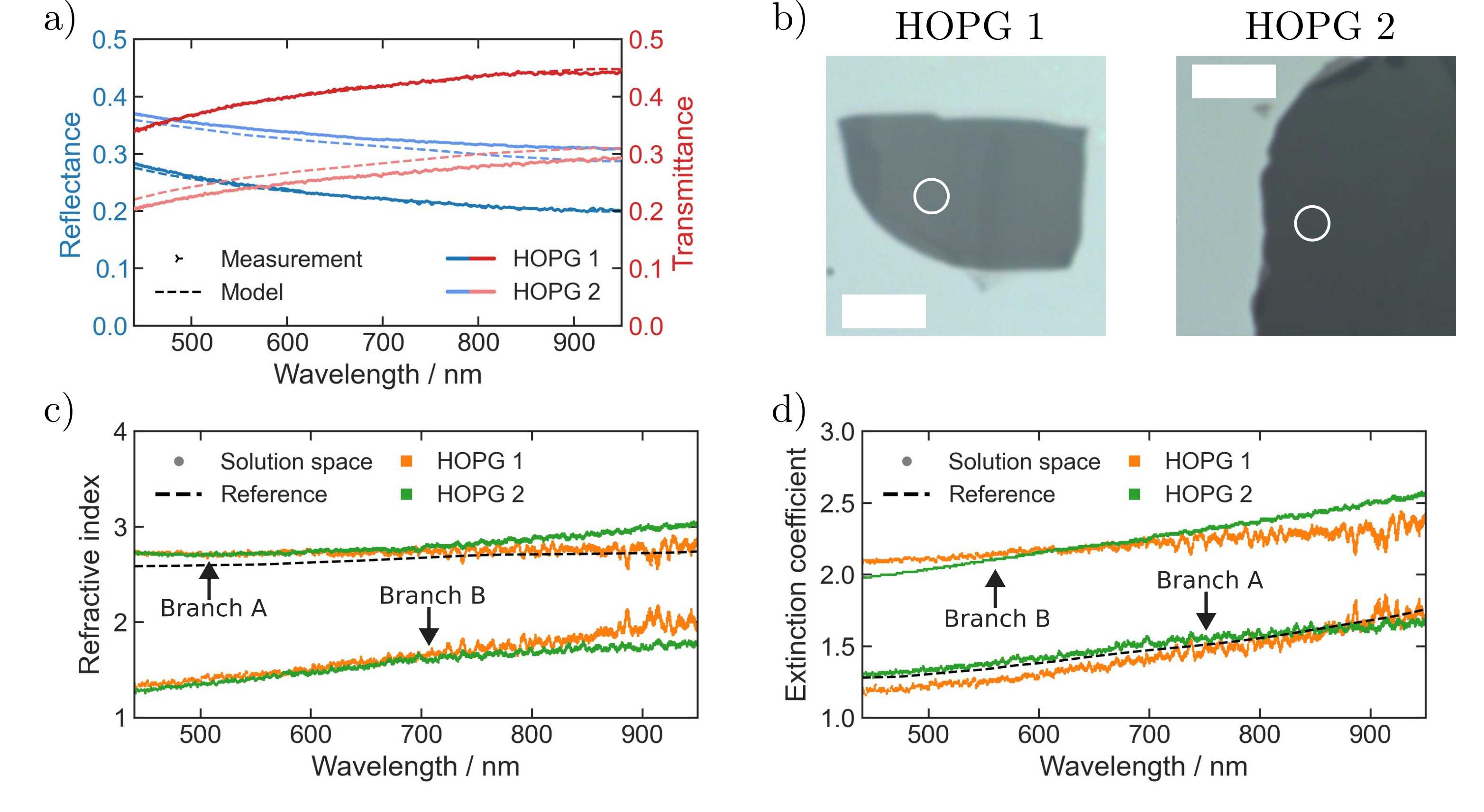}
\caption{(a) Measured and modeled reflectance ($R$) and transmittance ($T$) spectra of 2 \acs{HOPG} flakes on a \qty{500}{\micro\meter} thick \acs{BF33} glass substrate using the anisotropic reference dispersion from~\textcite{Boosalis.2015} for the 100x objective lens with effective \ac{NA} of 0.90 ($R$) and 0.60 ($T$). Micrographs of the samples with thicknesses of ca. \qty{17}{\nano\meter} (HOPG 1) and \qty{29}{\nano\meter} (HOPG 2) determined by \acs{AFM}. The scale bar represents \qty{5}{\micro\meter}. Extracted solution spaces of both flakes and the reference values by~\textcite{Boosalis.2015} for the (c) refractive indices $n$ and (d) extinction coefficients $k$.}
\label{Figure 3}
\end{figure}

Fig.~\ref{Figure 3} illustrates the results for 2 \ac{HOPG} flakes exfoliated on a \qty{500}{\micro\meter} thick \ac{BF33} substrate with thicknesses of \qty{17.0\pm1.7}{\nano\meter} (\ac{HOPG} 1) and \qty{28.5\pm2.7}{\nano\meter} (HOPG 2), determined via \ac{AFM}.
For a first impression, Fig.~\ref{Figure 3} presents the measured and modeled reflectance and transmittance spectra, where reference modeling was performed with the mean thickness from \ac{AFM} and the full anisotropic tensor of the complex refractive index via 4×4 \ac{TMM} modeling~\cite{Boosalis.2015,Schwarz.2023}. 
While the results for sample 1 demonstrate precise agreemeent between measurement and simulation, the model deviates significantly for sample 2.
A possible reason is the increased inhomogeneity of sample 2 symbolized by the heightened standard deviation.
Thus, assuming a thickness at the upper end of the AFM thickness range would result in significantly better agreement with the measurement data due to greater absorption.
Since these conclusions cannot be drawn when the refractive indices are unknown, the following calculations are nevertheless based on the mean thickness.
The corresponding micrographs of the flakes, with the position of the measurement spot with a diameter of \qty{2}{\micro\meter} indicated as a white circle, are shown in Fig.~\ref{Figure 3}b.

Figs.~\ref{Figure 3}c and d depict the extracted effective, i.e., isotropic refractive indices and extinction coefficients for both samples.
As these values are dominated by the in-plane contributions~\cite{Schwarz.2023,Ermolaev.2020}, they are compared with their in-plane counterparts of the anisotropic tensor applied for modeling in Fig.~\ref{Figure 3}a~\cite{Boosalis.2015}, which, in the case of \ac{HOPG}, is of uniaxial nature.
Here, the thresholds for the \ac{SErr} were \qty{1e-6}{} and the iteration of $n$ and $k$ was performed over the ranges of 1-4 and 0-3, respectively, with steps of 0.01.
It becomes clear, that for each of the two samples, there are two similar solution branches for $n$ and $k$.
The branch with the higher $n$, denoted as branch A, corresponds to the smaller $k$, and vice versa.
Such ambiguity constitutes a fundamental limitation of the method. 
In favorable cases, it may be mitigated or even resolved, for example by measuring a large number of samples or by using different \acp{NA} values. 
In the absence of \textit{a priori} knowledge of $n$ and/or $k$, however, the ambiguity cannot be resolved.

Another approach is to restrict the solution space based on \ac{DFT} simulations of the expected material parameters~\cite{Schue.2022,Gu.2023}. 
However, measurements at high \ac{NA} introduce an additional source of uncertainty due to the thickness-dependent influence of the out-of-plane component on the measurement spectra and the resulting thickness-dependent complex refractive indices. 
The influence of the out-of-plane anisotropy is especially relevant for \ac{HOPG}, as shown in our previous work~\cite{Schwarz.2023}.
In addition to the already amplified error in the anisotropic reference modeling shown in Fig.~\ref{Figure 3}a -- presumably caused by the inhomogeneity of sample 2 -- this thickness-dependent effect of uniaxial anisotropy on the measurement spectra also explains why the solution branches do not coincide with each other and with the reference values.
Nonetheless, the solutions for the isotropic approximation of the complex refractive indices of \ac{HOPG} yield a great estimate.
This is highlighted by the comparison of various in-plane complex refractive index for \ac{HOPG} and graphite from literature in Fig.~S11~\cite{Djurisic.1999,Jellison.2007,Kravets.2010,Boosalis.2015,Song.2018,Toksumakov.2026}.
We limited the comparison to solution branch A of \ac{HOPG} 1.
There, the result for \ac{HOPG} 1 fits well with the literature data, where wavelength-dependent differences of up to 0.6 occur between the individual extracted values for $n$ and $k$.

\begin{figure}[H]
\includegraphics[width=\linewidth]{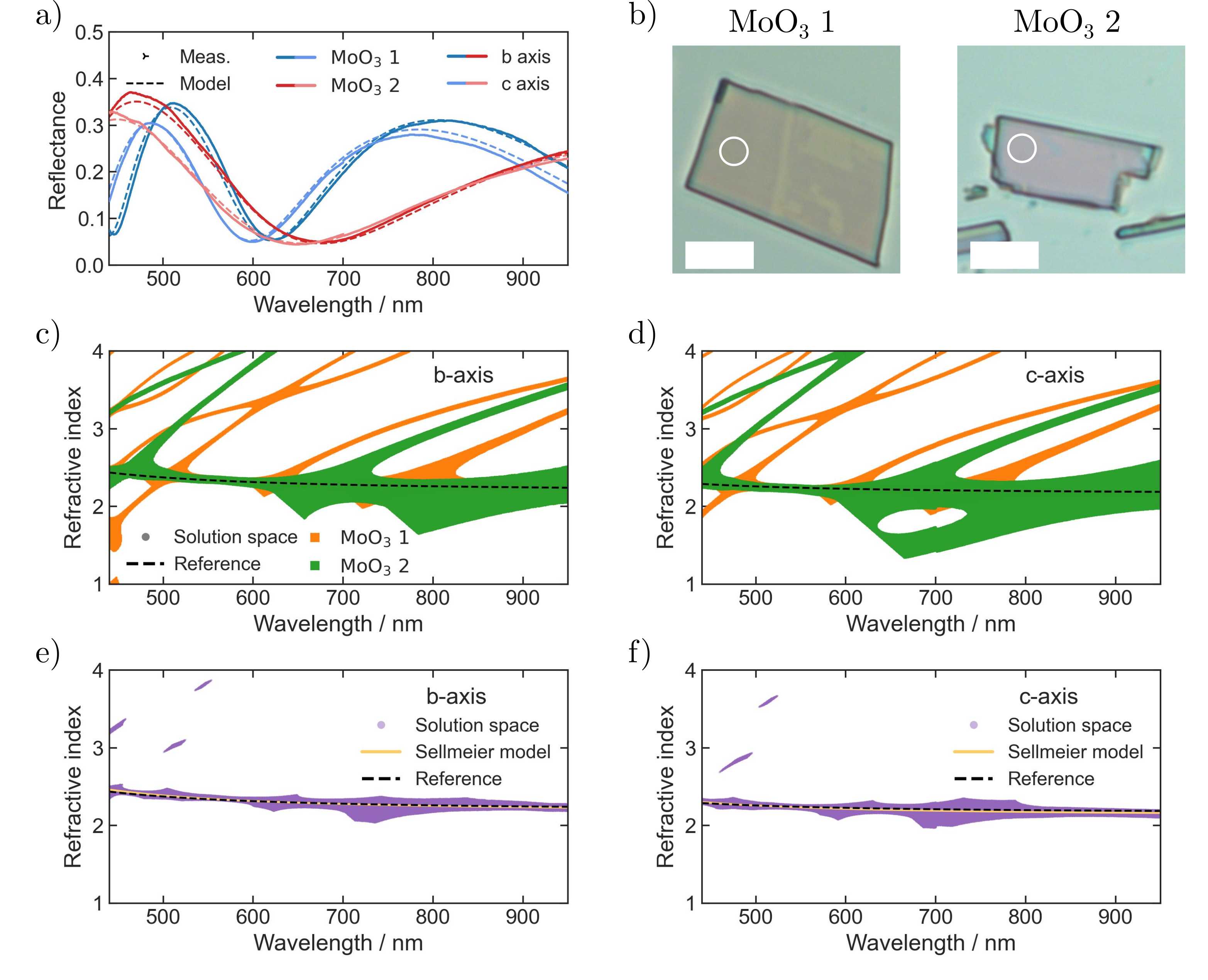}
\caption{(a) Measured and modeled linearly polarized reflectance spectra along the b- and c-axes of 2 \ce{MoO3} flakes on a \qty{500}{\micro\meter} thick \acs{BF33} glass substrate using the anisotropic reference dispersion from~\textcite{AndresPenares.2021} for the 100x objective lens with \ac{NA} 0.90. Micrographs of the samples with thicknesses of ca. \qty{279}{\nano\meter} (\ce{MoO3} 1) and \qty{154}{\nano\meter} (\ce{MoO3} 2) determined by \acs{AFM}. Extracted solution spaces for both samples and the reference values from~\textcite{AndresPenares.2021} for the refractive indices $n$ along the (c) b-axis and (d) c-axis. Combined solution space for both samples, Sellmeier model and the reference values by~\textcite{AndresPenares.2021} along the (e) b-axis and (f) c-axis.}
\label{Figure 4}
\end{figure}

Finally, the introduction of linearly polarized light enables the extraction of the refractive indices along the in-plane crystal axes of biaxial materials.
Since linearly polarized light parallel to the in-plane axes effectively reduces the biaxial behavior to that of a uniaxial material like \ac{HOPG}, the determined material properties are again, in principle, isotropic approximations showing thickness-dependent impact of the out-of-plane component.
Fig.~\ref{Figure 4}a presents the measured and modeled linearly polarized reflectance spectra of 2 \ce{MoO3} flakes exfoliated on a \qty{500}{\micro\meter} thick \ac{BF33} substrate along the in-plane crystal b- and c-axes.
The crystal orientation was identified by the rectangular cleavage characterisic of \ce{MoO3}, where the long edge is parallely aligned to the b-axis~\cite{AndresPenares.2021,Schwarz.2026}. 
The rectangular shape is evident in the micrographs of the \qty{279.1\pm4.8}{\nano\meter} (\ce{MoO3} 1) and \qty{153.7\pm3.2}{\nano\meter} (\ce{MoO3} 2) thick flakes, as determined by \ac{AFM}, in Fig.~\ref{Figure 4}b.
The modeling in Fig.~\ref{Figure 4}a was performed with the mean thicknesses and the full anisotropic tensor from literature~\cite{AndresPenares.2021} and 4×4 \ac{TMM}~\cite{Schwarz.2023,Schwarz.2026}.
Overall, a good agreement between the measured and modeled spectra was achieved.
However, the deviation is slightly increased for sample 1.
Due to the large bandgap of ca. \qty{3.4}{\electronvolt} ($\approx \qty{365}{\nano\meter}$), the extinction coefficient was assumed to be 0~\cite{AndresPenares.2021,AbediniDereshgi.2023}.
Hence, only polarized reflectance measurements are required to determine the refractive indices. 
According to that, the Figs~\ref{Figure 4}c and d depict the extracted solution spaces for the refractive indices along the b- and c-axes of \ce{MoO3} from the measured data in Fig.~\ref{Figure 4}a.
The \ce{SE} threshold was set to \qty{1e-5}{} and the iteration range from 1-4 was scanned in steps of 0.001.
For both axes, the main branch with the reference values is accompanied by side branches due to the thickness-dependent interference pattern.
While in the case of the \ce{Si_xN_y} film in Fig.~S9, the ambiguity was resolved through the overlap of solutions obtained with different objective lenses, we exploit the differences in the solution spaces of both flakes.
Consequently, Figs.~\ref{Figure 4}e and f show the overlap of the solutions along the b- and c-axes.
The numerical artifacts caused by the interference pattern are mostly eliminated and the remaining continuous solutions are the main branches containing the reference values.

When a Sellmeier model was fitted to each of these paths, the curves coincide almost perfectly with the reference dispersion.
The model parameters for the b-axis are: $\mathrm{B}=\qty{3.75539\pm0.01647}{}$ and $\mathrm{C}=\qty{0.05021\pm0.00105}{\micro\meter\squared}$.
Those along the c-axis are: $\mathrm{B}=\qty{3.53194 \pm 0.01769}{}$ and $\mathrm{C}=\qty{0.03472\pm0.00128}{\micro\meter\squared}$.
Compared to other literature values for the in-plane refractive indices of \ce{MoO3} in Fig.~S12~\cite{Lajaunie.2013,AbediniDereshgi.2023}, our results support the reference values from \cite{AndresPenares.2021}.
Moreover, there are enormous wavelength-dependent differences of up to 1 between different publications.

For \ce{MoO3}, the isotropic approximation shows a negligible deviation from the reference values, whereas there is a distinct difference for \ac{HOPG}.
The reason is the small in-plane/out-of-plane birefringence of 0.25 to 0.33 at \qty{632}{\nano\meter} for \ce{MoO3}~\cite{AndresPenares.2021}.
In contrast, \ce{HOPG} exhibits large birefringence $\Delta n$ and linear dichroism $\Delta k$ with values above 1 and, as a consequence, the influence of the out-of-plane is remarkably increased~\cite{Schwarz.2023,Boosalis.2015}.
Hence, the isotropic approximation of the complex refractive index is more strongly influenced by the out-of-plane component than in the case of \ce{MoO3}.

If additional linearly polarized transmittance spectra are recorded, it is possible to simultaneously determine $n$ and $k$ along the in-plane crystal axes of biaxial materials. 
A level of precision similar to that achieved for \ce{MoO3} is expected.
For biaxial materials, further independent physical quantities are obtainable when measurements with linearly polarized light between the crystal axes are recorded.
The utmost requirement for this, is, however, the exact knowledge of the relative position of the polarization angle to the crystal axes necessary for precise 4×4 \ac{TMM} modeling~\cite{Schwarz.2026}.   
Though, by including polarized measurements between the crystal axes, uncertainties due to polarization dependent components of the measurement setup may arise~\cite{Schwarz.2026}. 
Aside from that, the computational effort will increase considerably for the parallel fitting of spectra with multiple polarization angles.
In addition to the precise determination of the orientation of the crystallographic axes and the adjustment of the polarizer, potential sources of uncertainty include thickness inhomogeneities of the flakes, as evidenced by the comparatively large standard deviations obtained from the AFM measurements.

In principle, several approaches can be considered for the simultaneous extraction of the in-plane and out-of-plane complex refractive indices from microspectroscopic measurements. 
One possibility is based on the assumption that the influence of the out-of-plane component diminishes with decreasing flake thickness.
Accordingly, the refractive indices obtained from flakes with a low number of layers -- while still exhibiting bulk-like behavior at around ten layers depending on the specific material and dielectric tensor component~\cite{Li.2017b,Gu.2019,Chen.2023b} -- may be approximated as predominantly representing the in-plane optical properties. 
These values could then serve as input for anisotropic modeling of thicker flakes, where only the out-of-plane refractive indices would remain as unknown parameters.

An alternative approach is the measurement of flakes under varying \acp{NA}. 
By artificially reducing the \ac{NA}~\cite{Knockl.2025}, the influence of the out-of-plane component can be significantly suppressed even for high-magnifica\-tion objectives, owing to the resulting near-normal incidence. 
The refractive indices extracted under these conditions would therefore predominantly correspond to those parallel to the individual layers. 
Subsequent analysis of measurements acquired at high \ac{NA} would again enable anisotropic modeling.
The main limitations of this approach are the assumptions made and the reduced lateral resolution associated with decreasing the effective \ac{NA} at high magnification.

\section{Conclusion}

In summary, we demonstrate a reliable and non-destructive method for the determination of the complex refractive indices of samples with lateral dimensions on the micrometer scale via microspectroscopy.
The proposed approach combines an optical microscope with a spectrometer, enabling reflectance and transmittance measurements while maintaining a simple experimental setup.
We successfully transferred our objective lens dependent modeling from reflectance studies to transmittance, required due to the large \ac{NA} of high magnification objective lens and their strong influence on the measured spectra.
By doing so, we achieve similar precision for the thickness determination of a \ce{Si_xN_y} film from transmittance data, as previously demonstrated for reflectance.   

(Complex) refractive indices were extracted from the measured reflectance and transmittance data for a large number of samples by comparing them with Fresnel-based modeling without the need of further formalisms for the optical dispersion like the Drude-Lorentz model or the Kramers-Kronig relations.
Starting with thick transparent glasses characterized by their incoherent behavior, excellent agreement with refractive indices from literature was obtained for objective lenses with low and high \acp{NA}.
Moreover, the intricate extinction coefficients of a thick absorbing 4H-\ce{SiC} substrate were retrieved with high accuracy.
The ambiguous solutions for the dispersion of a \ce{Si_xN_y} film, due to the interference pattern and the numerical nature of the approach, could be resolved by superimposing the solution spaces for lenses with different \acp{NA}.
For small scale flakes of the van der Waals materials \ac{HOPG} and \ce{MoO3}, such ambiguity was eliminated by analyzing samples with varying thicknesses.
For \ce{MoO3}, as a biaxial material, the refractive indices were succesfully determined along the in-plane crystal axes through the utilization of linearly polarized light.

While simple Savitzky--Golay filtering of the extracted solution spaces generally provided good agreement with reference dispersions from literature, the application of a simple Sellmeier equation -- where applicable -- further improved the accuracy.
In addition, the limitations arising from the low sensitivity to the out-of-plane component in anisotropic samples and their isotropic approximation are discussed, and possible approaches for determining the full anisotropic tensor of the refractive index are proposed.
Thus, the method presents itself as a promising alternative to state-of-the-art methods for determining the material dispersion, such as \acl{SE} and its variations for micrometer scale characterization, with a comparatively simple setup.

\section*{CRediT authorship contribution statement}
\noindent
\textbf{Schwarz Julian:} Conceptualization, Data curation, Formal analysis, Investigation, Methodology, Software, Visualization, Writing - original draft, Writing - review and editing. \\
\textbf{Bauer Johannes:} Investigation, Software,  Writing - review and editing. \\ 
\textbf{Rommel Mathias:} Supervision, Writing - review and editing. \\
\textbf{Hutzler Andreas:} Supervision, Writing - review and editing.

\section*{Declaration of Competing Interest}
The authors declare that they have no known competing financial interests or personal relationships that could have appeared to influence the work reported in this article.

\section*{Data availability}
The data that support the findings of this study are available from the corresponding author upon reasonable request.

\section*{Funding}
This research did not receive any specific grant from funding agencies in the public, commercial, or not-for-profit sectors.

\printbibliography

@article{AbediniDereshgi.2023,
 
 author = {{Abedini Dereshgi}, Sina and Lee, Yea-Shine and Larciprete, Maria Cristina and Centini, Marco and Dravid, Vinayak P. and Aydin, Koray},
 year = {2023},
 title = {Low-Symmetry {$\alpha$}-\ce{MoO3} Heterostructures for Wave Plate Applications in Visible Frequencies},
 pages = {2202603},
 volume = {11},
 number = {7},
 issn = {2195-1071},
 journal = {Advanced Optical Materials},
 doi = {10.1002/adom.202202603},

}

@article{Wu.2024,
 author = {Wu, Wanglong and Liu, Zhiyuan and Qiu, Zhicong and Wu, Ziqiao and Li, Zhongming and Yang, Xiong and Han, Lixiang and Li, Chengyuan and Huo, Nengjie and Wang, Xiaozhou and Yao, Jiandong and Zheng, Zhaoqiang and Li, Jingbo},
 year = {2024},
 title = {An Ultrasensitive \ce{ReSe2}/\ce{WSe2} Heterojunction Photodetector Enabled by Gate Modulation and its Development in Polarization State Identification},
 pages = {2301410},
 volume = {12},
 number = {2},
 issn = {2195-1071},
 journal = {Advanced Optical Materials},
 doi = {10.1002/adom.202301410}
}

@misc{AF32eco,
 author = {{SCHOTT AG}},
 title = {AF 32{\circledR} eco Thin Glass - Data sheet},
 url = {https://refractiveindex.info/?shelf=specs&book=SCHOTT-misc&page=AF32ECO},
 note = {Accessed 2026-07-01},
}

@article{AndresPenares.2021,

 author = {Andres-Penares, Daniel and Brotons-Gisbert, Mauro and Bonato, Cristian and S{\'a}nchez-Royo, Juan F. and Gerardot, Brian D.},
 year = {2021},
 title = {Optical and dielectric properties of \ce{MoO3} nanosheets for van der Waals heterostructures},
 pages = {223104},
 volume = {119},
 number = {22},
 issn = {0003-6951},
 journal = {Applied Physics Letters},
 doi = {10.1063/5.0066219},
}

@article{BakerFinch.2011,
 author = {Baker-Finch, Simeon C. and McIntosh, Keith R.},
 year = {2011},
 title = {Reflection of normally incident light from silicon solar cells with pyramidal texture},
 pages = {406--416},
 volume = {19},
 number = {4},
 issn = {1062-7995},
 journal = {Progress in Photovoltaics: Research and Applications},
 doi = {10.1002/pip.1050}
}

@article{Beliaev.2022,
 author = {Beliaev, Leonid Yu. and Shkondin, Evgeniy and Lavrinenko, Andrei V. and Takayama, Osamu},
 year = {2022},
 title = {Optical, structural and composition properties of silicon nitride films deposited by reactive radio-frequency sputtering, low pressure and plasma-enhanced chemical vapor deposition},
 pages = {139568},
 volume = {763},
 issn = {0040-6090},
 journal = {Thin Solid Films},
 doi = {10.1016/j.tsf.2022.139568}
}

@misc{BF33,
 author = {{SCHOTT AG}},
 title = {BOROFLOAT{\circledR} 33 -- Data sheet},
 url = {https://media.schott.com/api/public/content/cda43a92330145c9b34db0373098ec32?v=87d55030&download=true},
 note = {Accessed 2026-07-01}
}

@phdthesis{Boosalis.2015,
 author = {Boosalis, Alexander},
 year = {2015},
 title = {Ellipsometric Characterization of Silicon and Carbon Junctions for Advanced Electronics},
 school = {{University of Nebraska-Lincoln}},
 type = {Ph.D. Dissertation}
}

@article{Chu.2020,
 author = {Chu, Yuhui and Zhang, Ziyang},
 year = {2020},
 title = {Birefringent and Complex Dielectric Functions of Monolayer \ce{WSe2} Derived by Spectroscopic Ellipsometer},
 pages = {12665--12671},
 volume = {124},
 number = {23},
 issn = {1932-7447},
 journal = {The Journal of Physical Chemistry C},
 doi = {10.1021/acs.jpcc.0c03691}
}

@article{Djurisic.1999,
 author = {Djuri{\v{s}}i{\'c}, Aleksandra B. and Li, E. Herbert},
 year = {1999},
 title = {Optical properties of graphite},
 pages = {7404--7410},
 volume = {85},
 number = {10},
 issn = {0021-8979},
 journal = {Journal of Applied Physics},
 doi = {10.1063/1.369370}
}

@article{Duttagupta.2012,
 author = {Duttagupta, Shubham and Ma, Fajun and Hoex, Bram and Mueller, Thomas and Aberle, Armin G.},
 year = {2012},
 title = {Optimised Antireflection Coatings using Silicon Nitride on Textured Silicon Surfaces based on Measurements and Multidimensional Modelling},
 pages = {78--83},
 volume = {15},
 issn = {1876-6102},
 journal = {Energy Procedia},
 doi = {10.1016/j.egypro.2012.02.009}
}

@article{Gu.2023,
 author = {Gu, H. and Guo, Zhengfeng and Huang, Liusheng and Fang, Mingsheng and Liu, Shiyuan},
 year = {2023},
 title = {Investigations of Optical Functions and Optical Transitions of 2D Semiconductors by Spectroscopic Ellipsometry and DFT},
 volume = {13},
 number = {1},
 issn = {2079-4991},
 journal = {Nanomaterials},
 doi = {10.3390/nano13010196}
}

@article{Schue.2022,
 author = {Schu{\'e}, L{\'e}onard and Goudreault, F{\'e}lix A. and Righi, Ariete and Resende, Geovani C. and Lefebvre, Val{\'e}rie and Godbout, {\'E}mile and Tie, Monique and Ribeiro, Henrique B. and Heinz, Tony F. and Pimenta, Marcos A. and C{\^o}t{\'e}, Michel and Franc{\oe}ur, S{\'e}bastien and Martel, Richard},
 year = {2022},
 title = {Visible Out-of-plane Polarized Luminescence and Electronic Resonance in Black Phosphorus},
 pages = {2851--2858},
 volume = {22},
 number = {7},
 issn = {1530-6984},
 journal = {Nano Letters},
 doi = {10.1021/acs.nanolett.1c04998}
}

@article{Knockl.2025,
 author = {Kn{\"o}ckl, Ernst and Bernard, Alexandre and Holleitner, Alexander and Kastl, Christoph},
 year = {2025},
 title = {Polarized optical contrast spectroscopy of in plane anisotropic van der Waals materials},
 pages = {15344},
 volume = {15},
 number = {1},
 issn = {2045-2322},
 journal = {Scientific Reports},
 doi = {10.1038/s41598-025-96894-8}
}

@article{Chen.2023b,
 author = {Chen, Chengxiang and Wang, Zhenyu and Zhang, Bo and Zhang, Zixuan and Zhang, Jinying and Cheng, Yonghong and Wu, Kai and Zhou, Jun},
 year = {2023},
 title = {Thickness-Dependent Dielectric Screening in Few-Layer Phosphorus},
 pages = {4962--4969},
 volume = {14},
 number = {21},
 issn = {1948-7185},
 journal = {The Journal of Physical Chemistry Letters},
 doi = {10.1021/acs.jpclett.3c00608}
}

@book{OpticalSocietyofAmerica.1995b,
 year = {1995},
 title = {Handbook of Optics},
 edition = {2. Ed.},
 volume = {2},
 publisher = {McGraw-Hill},
 isbn = {9780070479746},
 editor = {Bass, Michael}
}

@article{Li.2017b,

 author = {Li, Xiao-Li and Han, Wen-Peng and Wu, Jiang-Bin and Qiao, Xiao-Fen and Zhang, Jun and Tan, Ping-Heng},
 year = {2017},
 title = {Layer-Number Dependent Optical Properties of 2D Materials and Their Application for Thickness Determination},
 pages = {1604468},
 volume = {27},
 number = {19},
 issn = {1616-301X},
 journal = {Advanced Functional Materials},
 doi = {10.1002/adfm.201604468}
}

@article{EnricoNichelatti.2002,
 author = {Nichelatti, Enrico},
 year = {2002},
 title = {Complex refractive index of a slab from reflectance and transmittance: analytical solution},
 pages = {400},
 volume = {4},
 number = {4},
 issn = {1464-4258},
 journal = {Journal of Optics A: Pure and Applied Optics},
 doi = {10.1088/1464-4258/4/4/306}
}

@article{Ermolaev.2021,
 author = {Ermolaev, G. A. and Grudinin, D. V. and Stebunov, Y. V. and Voronin, K. V. and Kravets, V. G. and Duan, J. and Mazitov, A. B. and Tselikov, G. I. and Bylinkin, A. and Yakubovsky, D. I. and Novikov, S. M. and Baranov, D. G. and Nikitin, A. Y. and Kruglov, I. A. and Shegai, T. and Alonso-Gonz{\'a}lez, P. and Grigorenko, A. N. and Arsenin, A. V. and Novoselov, K. S. and Volkov, V. S.},
 year = {2021},
 title = {Giant optical anisotropy in transition metal dichalcogenides for next-generation photonics},
 pages = {854},
 volume = {12},
 number = {1},
 issn = {2041-1723},
 journal = {Nature Communications},
 doi = {10.1038/s41467-021-21139-x}
}

@article{Gu.2019,
 author = {Gu, H. and Song, Baokun and Fang, Mingsheng and Hong, Yilun and Chen, Xiuguo and Jiang, Hao and Ren, Wencai and Liu, Shiyuan},
 year = {2019},
 title = {Layer-dependent dielectric and optical properties of centimeter-scale 2D \ce{WSe2}: evolution from a single layer to few layers},
 pages = {22762--22771},
 volume = {11},
 number = {47},
 issn = {2040-3364},
 journal = {Nanoscale},
 doi = {10.1039/C9NR04270A}
}

@article{Guo.2024,
 author = {Guo, Q. and Zhang, Qiuhong and Zhang, Tan and Zhou, Jun and Xiao, Shumin and Wang, Shijie and Feng, Yuan Ping and Qiu, Cheng-Wei},
 year = {2024},
 title = {Colossal in-plane optical anisotropy in a two-dimensional van der Waals crystal},
 pages = {1170--1175},
 volume = {18},
 number = {11},
 issn = {1749-4885},
 journal = {Nature Photonics},
 doi = {10.1038/s41566-024-01501-3}
}

@article{Hsu.2019,
 author = {Hsu, Chunwei and Frisenda, Riccardo and Schmidt, Robert and Arora, Ashish and de Vasconcellos, Steffen Michaelis and Bratschitsch, Rudolf and {van der Zant}, Herre S. J. and Castellanos-Gomez, Andres},
 year = {2019},
 title = {Thickness-Dependent Refractive Index of 1L, 2L, and 3L \ce{MoS2}, \ce{MoSe2}, \ce{WS2}, and \ce{WSe2}},
 pages = {1900239},
 volume = {7},
 number = {13},
 issn = {2195-1071},
 journal = {Advanced Optical Materials},
 doi = {10.1002/adom.201900239}
}

@article{Hutzler.2017,
 author = {Hutzler, A. and Matthus, C. D. and Rommel, M. and Frey, L.},
 year = {2017},
 title = {Generalized approach to design multi-layer stacks for enhanced optical detectability of ultrathin layers},
 pages = {021909},
 volume = {110},
 number = {2},
 issn = {0003-6951},
 journal = {Applied Physics Letters},
 doi = {10.1063/1.4973968}
}

@article{Hutzler.2019,

 author = {Hutzler, Andreas and Matthus, Christian D. and Dolle, Christian and Rommel, Mathias and Jank, Michael P. M. and Spiecker, Erdmann and Frey, Lothar},
 year = {2019},
 title = {Large-Area Layer Counting of Two-Dimensional Materials Evaluating the Wavelength Shift in Visible-Reflectance Spectroscopy},
 pages = {9192--9201},
 volume = {123},
 number = {14},
 issn = {1932-7447},
 journal = {The Journal of Physical Chemistry C},
 doi = {10.1021/acs.jpcc.9b00957}
}

@article{Hutzler.2020,
 author = {Hutzler, Andreas and Fritsch, Birk and Matthus, Christian D. and Jank, Michael P. M. and Rommel, Mathias},
 year = {2020},
 title = {Highly accurate determination of heterogeneously stacked Van-der-Waals materials by optical microspectroscopy},
 pages = {13676},
 volume = {10},
 number = {1},
 issn = {2045-2322},
 journal = {Scientific Reports},
 doi = {10.1038/s41598-020-70580-3}
}

@article{Jellison.2007,
 author = {Jellison, G. E. and Hunn, J. D. and Lee, Ho Nyung},
 year = {2007},
 title = {Measurement of optical functions of highly oriented pyrolytic graphite in the visible},
 pages = {085125},
 volume = {76},
 number = {8},
 issn = {1098-0121},
 journal = {Physical Review B},
 doi = {10.1103/PhysRevB.76.085125}
}

@article{Kenaz.2023,
 author = {Kenaz, Ralfy and Rapaport, Ronen},
 year = {2023},
 title = {Mapping spectroscopic micro-ellipsometry with sub-5 microns lateral resolution and simultaneous broadband acquisition at multiple angles},
 pages = {023908},
 volume = {94},
 number = {2},
 issn = {0034-6748},
 journal = {Review of Scientific Instruments},
 doi = {10.1063/5.0123249}
}

@article{Kravets.2010,
 author = {Kravets, V. G. and Grigorenko, A. N. and Nair, R. R. and Blake, P. and Anissimova, S. and Novoselov, K. S. and Geim, A. K.},
 year = {2010},
 title = {Spectroscopic ellipsometry of graphene and an exciton-shifted van Hove peak in absorption},
 pages = {155413},
 volume = {81},
 number = {15},
 issn = {1098-0121},
 journal = {Physical Review B},
 doi = {10.1103/PhysRevB.81.155413}
}

@article{Lajaunie.2013,
 author = {Lajaunie, L. and Boucher, F. and Dessapt, R. and Moreau, P.},
 year = {2013},
 title = {Strong anisotropic influence of local-field effects on the dielectric response of {$\alpha$}-\ce{MoO3}},
 pages = {115141},
 volume = {88},
 number = {11},
 issn = {1098-0121},
 journal = {Physical Review B},
 doi = {10.1103/PhysRevB.88.115141}
}

@article{Lee.2019,
 author = {Lee, Seong--Yeon and Jeong, Tae--Young and Jung, Suyong and Yee, Ki--Ju},
 year = {2019},
 title = {Refractive Index Dispersion of Hexagonal Boron Nitride in the Visible and Near--Infrared},
 pages = {1800417},
 volume = {256},
 number = {6},
 issn = {0370-1972},
 journal = {Physica Status Solidi B},
 doi = {10.1002/pssb.201800417}
}

@article{Lee.2021b,
 author = {Lee, Seong-Yeon and Yee, Ki-Ju},
 year = {2021},
 title = {Black phosphorus phase retarder based on anisotropic refractive index dispersion},
 pages = {015020},
 volume = {9},
 number = {1},
 issn = {2053-1583},
 journal = {2D Materials},
 doi = {10.1088/2053-1583/ac3a99}
}

@article{Munkhbat.2022,
 author = {Munkhbat, Battulga and Wr{\'o}bel, Piotr and Antosiewicz, Tomasz J. and Shegai, Timur O.},
 year = {2022},
 title = {Optical Constants of Several Multilayer Transition Metal Dichalcogenides Measured by Spectroscopic Ellipsometry in the 300--1700 nm Range: High Index, Anisotropy, and Hyperbolicity},
 pages = {2398--2407},
 volume = {9},
 number = {7},
 issn = {2330-4022},
 journal = {ACS Photonics},
 doi = {10.1021/acsphotonics.2c00433}
}

@article{Ross.2020,
 author = {Ross, Aaron M. and Patern{\`o}, Giuseppe M. and {Dal Conte}, Stefano and Scotognella, Francesco and Cinquanta, Eugenio},
 year = {2020},
 title = {Anisotropic Complex Refractive Indices of Atomically Thin Materials: Determination of the Optical Constants of Few-Layer Black Phosphorus},
 volume = {13},
 number = {24},
 issn = {1996-1944},
 journal = {Materials},
 doi = {10.3390/ma13245736}
}

@article{Schwarz.2023,
 author = {Schwarz, Julian and Niebauer, Michael and Kole{\'s}nik-Gray, Maria and Szabo, Maximilian and Baier, Leander and Chava, Phanish and Erbe, Artur and Krsti{\'c}, Vojislav and Rommel, Mathias and Hutzler, Andreas},
 year = {2023},
 title = {Correlating Optical Microspectroscopy with 4$\times$4 Transfer Matrix Modeling for Characterizing Birefringent Van der Waals Materials},
 pages = {2300618},
 volume = {7},
 number = {10},
 issn = {2366-9608},
 journal = {Small Methods},
 doi = {10.1002/smtd.202300618}
}

@article{Schwarz.2025,
 author = {Schwarz, Julian and Niebauer, Michael and R{\"o}mling, Lukas and Pham, Adrian and Kole{\'s}nik-Gray, Maria and Evanschitzky, Peter and Vogel, Nicolas and Krsti{\'c}, Vojislav and Rommel, Mathias and Hutzler, Andreas},
 year = {2025},
 title = {Spectro-Spatial Unmixing in Optical Microspectroscopy for Thickness Determination of Layered Materials},
 pages = {2402502},
 volume = {13},
 number = {5},
 issn = {2195-1071},
 journal = {Advanced Optical Materials},
 doi = {10.1002/adom.202402502}
}

@article{Schwarz.2025b,
 author = {Schwarz, Julian and Dick, Jan and Beuer, Susanne and Rommel, Mathias and Hutzler, Andreas},
 year = {2025},
 title = {Modeling the partially detected backside reflectance of transparent substrates in reflectance microspectroscopy},
 pages = {103878},
 volume = {198},
 issn = {0968-4328},
 journal = {Micron},
 doi = {10.1016/j.micron.2025.103878}
}

@article{Schwarz.2026,
 author = {Schwarz, Julian and Bauer, Johannes and Ghazal, Roua and Schrotz, Albert-Marcel and Rommel, Mathias and Hutzler, Andreas},
 year = {2026},
 title = {Probing crystal axis orientation of birefringent materials via polarized microspectroscopy and anisotropic optical modeling},
 pages = {015029},
 volume = {8},
 number = {1},
 issn = {2515-7647},
 journal = {Journal of Physics: Photonics},
 doi = {10.1088/2515-7647/ae2e68}
}

@article{Slavich.2024c,
 author = {Slavich, A. and Ermolaev, G. and Zavidovskiy, I. and Grudinin, D. and Tatmyshevskiy, M. and Toksumakov, A. and Syuy, A. and Vyshnevyy, A. and Yakubovsky, D. and Novikov, S. and Ghazaryan, D. and Arsenin, A. and Volkov, V.},
 year = {2024},
 title = {Optical Properties of Biaxial van der Waals Crystals for Photonic Applications},
 pages = {S433-S438},
 volume = {88},
 number = {3},
 issn = {1062-8738},
 journal = {Bulletin of the Russian Academy of Sciences: Physics},
 doi = {10.1134/S1062873824709978}
}

@article{Song.2018,
 author = {Song, Baokun and Gu, Honggang and Zhu, Simin and Jiang, Hao and Chen, Xiuguo and Zhang, Chuanwei and Liu, Shiyuan},
 year = {2018},
 title = {Broadband optical properties of graphene and HOPG investigated by spectroscopic Mueller matrix ellipsometry},
 pages = {1079--1087},
 volume = {439},
 issn = {0169-4332},
 journal = {Applied Surface Science},
 doi = {10.1016/j.apsusc.2018.01.051}
}

@article{Toksumakov.2026,
 author = {Toksumakov, Adilet N. and Ermolaev, Georgy A. and Grudinin, Dmitriy V. and Slavich, Aleksandr S. and Pak, Nikolay V. and Tikhonowski, Gleb V. and Minnekhanov, Anton A. and Vyshnevyy, Andrey A. and Ruoff, Rodney S. and Tselikov, Gleb I. and Arsenin, Aleksey V. and Volkov, Valentyn S.},
 year = {2026},
 title = {Anisotropic dielectric function of graphite probed by far- and near-field spectroscopies},
 pages = {141902},
 volume = {128},
 number = {14},
 issn = {0003-6951},
 journal = {Applied Physics Letters},
 doi = {10.1063/5.0320505}
}

@phdthesis{Vogt.2015,
 author = {Vogt, Malte Ruben},
 year = {2015},
 title = {Development of Physical Models for the Simulation of Optical Properties of Solar Cell Modules},
 school = {{Leibniz Universit{\"a}t Hannover}},
 type = {Ph.D. Dissertation}
}

@article{Wan.2013,

 author = {Wan, Yimao and McIntosh, Keith R. and Thomson, Andrew F.},
 year = {2013},
 title = {Characterisation and optimisation of PECVD \ce{SiN_x} as an antireflection coating and passivation layer for silicon solar cells},
 pages = {032113},
 volume = {3},
 number = {3},
 issn = {2158-3226},
 journal = {AIP Advances},
 doi = {10.1063/1.4795108}
}

@book{Weber.1986,
 author = {Weber, M. J.},
 year = {1986},
 title = {CRC handbook of laser science and technology},
 publisher = {CRC Press Inc},
 volume = {4. Optical materials, Part 2 - Properties},
 isbn = {9780849335044}
}

@article{Yin.2013,
 author = {Yin, G. and Merschjann, C. and Schmid, M.},
 year = {2013},
 title = {The effect of surface roughness on the determination of optical constants of \ce{CuInSe2} and \ce{CuGaSe2} thin films},
 pages = {213510},
 volume = {113},
 number = {21},
 issn = {0021-8979},
 journal = {Journal of Applied Physics},
 doi = {10.1063/1.4809550}
}

@misc{Khadivianazar.2019,
 author = {Khadivianazar, Saba and Kole{\'s}nik-Gray, Maria and Krsti{\'c}, Vojislav and Weing{\"a}rtner, Roland and Kallinger, Birgit and Rommel, Mathias},
 title = {Doping Dependence of Optical Constants for n-Type (N) {4H}-\ce{SiC} Substrates},
 note ={Poster. ICSCRM, 18th International Conference on Silicon Carbide \& Related Materials, September 29 - October 4, 2019, Kyoto, Japan, (unpublished)}
}

@article{Frisenda.2017,
 author = {Frisenda, Riccardo and Niu, Yue and Gant, Patricia and Molina-Mendoza, Aday J. and Schmidt, Robert and Bratschitsch, Rudolf and Liu, Jinxin and Fu, Lei and Dumcenco, Dumitru and Kis, Andras and de Lara, David Perez and Castellanos-Gomez, Andres},
 year = {2017},
 title = {Micro-reflectance and transmittance spectroscopy: a versatile and powerful tool to characterize {2D} materials},
 pages = {074002},
 volume = {50},
 number = {7},
 issn = {0022-3727},
 journal = {Journal of Physics D: Applied Physics},
 doi = {10.1088/1361-6463/aa5256}
}

@article{Bing.2018,
 author = {Bing, Dan and Wang, Yingying and Bai, Jing and Du, Ruxia and Wu, Guoqing and Liu, Liyan},
 year = {2018},
 title = {Optical contrast for identifying the thickness of two-dimensional materials},
 pages = {128--138},
 volume = {406},
 issn = {0030-4018},
 journal = {Optics Communications},
 doi = {10.1016/j.optcom.2017.06.012}
}

@article{Niu.2018,
 author = {Niu, Yue and Gonzalez-Abad, Sergio and Frisenda, Riccardo and Marauhn, Philipp and Dr{\"u}ppel, Matthias and Gant, Patricia and Schmidt, Robert and Taghavi, Najme S. and Barcons, David and Molina-Mendoza, Aday J. and de Vasconcellos, Steffen M. and Bratschitsch, Rudolf and {Perez De Lara}, David and Rohlfing, Michael and Castellanos-Gomez, Andres},
 year = {2018},
 title = {Thickness-Dependent Differential Reflectance Spectra of Monolayer and Few-Layer \ce{MoS2}, \ce{MoSe2}, \ce{WS2} and \ce{WSe2}},
 volume = {8},
 number = {9},
 issn = {2079-4991},
 journal = {Nanomaterials},
 doi = {10.3390/nano8090725}
}

@article{Zhao.2020,

 author = {Zhao, Qinghua and Puebla, Sergio and Zhang, Wenliang and Wang, Tao and Frisenda, Riccardo and Castellanos-Gomez, Andres},
 year = {2020},
 title = {Thickness Identification of Thin \ce{InSe} by Optical Microscopy Methods},
 pages = {2000025},
 volume = {1},
 number = {2},
 issn = {2699-9293},
 journal = {Advanced Photonics Research},
 doi = {10.1002/adpr.202000025}
}

@article{Chaves.2020,
 author = {Chaves, A. and Azadani, J. G. and Alsalman, Hussain and {da Costa}, D. R. and Frisenda, R. and Chaves, A. J. and Song, Seung Hyun and Kim, Y. D. and He, Daowei and Zhou, Jiadong and Castellanos-Gomez, A. and Peeters, F. M. and Liu, Zheng and Hinkle, C. L. and Oh, Sang-Hyun and Ye, Peide D. and Koester, Steven J. and Lee, Young Hee and Avouris, Ph. and Wang, Xinran and Low, Tony},
 year = {2020},
 title = {Bandgap engineering of two-dimensional semiconductor materials},
 pages = {29},
 volume = {4},
 number = {1},
 issn = {2397-7132},
 journal = {Npj 2D Materials and Applications},
 doi = {10.1038/s41699-020-00162-4}
}

@book{Heavens.1991,
 author = {Heavens, Oliver S.},
 year = {1991},
 title = {Optical properties of thin solid films},
 edition = {2. Ed.},
 publisher = {Dover Publications},
 isbn = {0486669246},
}

@article{Savitzky.1964,
 author = {Savitzky, Abraham. and Golay, M. J. E.},
 year = {1964},
 title = {Smoothing and Differentiation of Data by Simplified Least Squares Procedures},
 pages = {1627--1639},
 volume = {36},
 number = {8},
 issn = {0003-2700},
 journal = {Analytical Chemistry},
 doi = {10.1021/ac60214a047}
}

@article{Everall.2010,

 author = {Everall, Neil J.},
 year = {2010},
 title = {Confocal Raman microscopy: common errors and artefacts},
 pages = {2512--2522},
 volume = {135},
 number = {10},
 issn = {0003-2654},
 journal = {Analyst},
 doi = {10.1039/C0AN00371A}
}

@article{Firsov.2016,
 author = {Firsov, D. D. and Komkov, O. S. and Fadeev, A. Yu and Lebedev, A. O.},
 year = {2016},
 title = {Evaluation of nitrogen incorporation into bulk 4H-SiC grown on seeds of different orientation from optical absorption spectra},
 pages = {012043},
 volume = {741},
 number = {1},
 issn = {1742-6588},
 journal = {Journal of Physics: Conference Series},
 doi = {10.1088/1742-6596/741/1/012043}
}

@article{Sellmeier.1872,
 author = {Sellmeier, W.},
 year = {1872},
 title = {{Ü}ber die durch die {Ä}therschwingungen erregten {M}itschwingungen der {K}örpertheilchen und deren {R}ückwirkung auf die {E}rsteren, besonders zur {E}rklärung der {D}ispersion und ihrer {A}nomalien},
 pages = {386--403},
 volume = {223},
 number = {11},
 issn = {0003-4169},
 journal = {Annales de Physique},
 doi = {10.1002/andp.18722231105}
}

@article{Zhang.2015,

 author = {Zhang, Hui and Ma, Yaoguang and Wan, Yi and Rong, Xin and Xie, Ziang and Wang, Wei and Dai, Lun},
 year = {2015},
 title = {Measuring the Refractive Index of Highly Crystalline Monolayer \ce{MoS2} with High Confidence},
 pages = {8440},
 volume = {5},
 number = {1},
 issn = {2045-2322},
 journal = {Scientific Reports},
 doi = {10.1038/srep08440}
}

@article{Green.2008,
 author = {Green, Martin A.},
 year = {2008},
 title = {Self-consistent optical parameters of intrinsic silicon at 300K including temperature coefficients},
 pages = {1305--1310},
 volume = {92},
 number = {11},
 issn = {0927-0248},
 journal = {Solar Energy Materials and Solar Cells},
 doi = {10.1016/j.solmat.2008.06.009}
}

@article{Peck.1972,
 author = {Peck, Edson R. and Reeder, Kaye},
 year = {1972},
 title = {Dispersion of Air},
 pages = {958--962},
 volume = {62},
 number = {8},
 issn = {1084-7529},
 journal = {Journal of the Optical Society of America A},
 doi = {10.1364/JOSA.62.000958}
}

@article{Li.2019,
 author = {Li, X. and Yu, Zhuoqing and Xiong, Xiong and Li, Tiaoyang and Gao, Tingting and Wang, Runsheng and Huang, Ru and Wu, Yanqing},
 year = {2019},
 title = {High-speed black phosphorus field-effect transistors approaching ballistic limit},
 pages = {eaau3194},
 volume = {5},
 number = {6},
 issn = {2375-2548},
 journal = {Science Advances},
 doi = {10.1126/sciadv.aau3194}
}

@article{Jin.2023,
 author = {Jin, Hyeok Jun and Park, Cheolmin and Byun, Hyo Hoon and Park, Seo Hak and Choi, Sung-Yool},
 year = {2023},
 title = {Electrically Modulated Single/Multicolor High Responsivity {2D} \ce{MoTe2}/\ce{MoS2} Photodetector for Broadband Detection},
 pages = {3027--3034},
 volume = {10},
 number = {9},
 issn = {2330-4022},
 journal = {ACS Photonics},
 doi = {10.1021/acsphotonics.3c00143}
}

@article{Dodda.2022,
 author = {Dodda, Akhil and Jayachandran, Darsith and Pannone, Andrew and Trainor, Nicholas and Stepanoff, Sergei P. and Steves, Megan A. and Radhakrishnan, Shiva Subbulakshmi and Bachu, Saiphaneendra and Ordonez, Claudio W. and Shallenberger, Jeffrey R. and Redwing, Joan M. and Knappenberger, Kenneth L. and Wolfe, Douglas E. and Das, Saptarshi},
 year = {2022},
 title = {Active pixel sensor matrix based on monolayer \ce{MoS2} phototransistor array},
 pages = {1379--1387},
 volume = {21},
 number = {12},
 issn = {1476-1122},
 journal = {Nature Materials},
 doi = {10.1038/s41563-022-01398-9}
}

@article{Li.2022c,
 author = {Li, Hao and Lin, Der-Yuh and {Di Renzo}, Anna and Puebla, Sergio and Frisenda, Riccardo and Gan, Xuetao and Quereda, Jorge and Xie, Yong and Al-Enizi, Abdullah M. and Nafady, Ayman and Castellanos-Gomez, Andres},
 year = {2022},
 title = {Stretching \ce{ReS2} along different crystal directions: Anisotropic tuning of the vibrational and optical responses},
 volume = {120},
 number = {6},
 issn = {0003-6951},
 journal = {Applied Physics Letters},
 doi = {10.1063/5.0081127}
}

@article{Wang.2023,
 author = {Wang, Junqi and Liu, Wei and Wang, Chunqing},
 year = {2023},
 title = {High-Performance Black Phosphorus Field-Effect Transistors with Controllable Channel Orientation},
 pages = {2201126},
 volume = {9},
 number = {3},
 issn = {2199160X},
 journal = {Advanced Electronic Materials},
 doi = {10.1002/aelm.202201126}
}

@article{Ermolaev.2020,
 author = {Ermolaev, Georgy A. and Stebunov, Yury V. and Vyshnevyy, Andrey A. and Tatarkin, Dmitry E. and Yakubovsky, Dmitry I. and Novikov, Sergey M. and Baranov, Denis G. and Shegai, Timur and Nikitin, Alexey Y. and Arsenin, Aleksey V. and Volkov, Valentyn S.},
 year = {2020},
 title = {Broadband optical properties of monolayer and bulk \ce{MoS2}},
 pages = {21},
 volume = {4},
 number = {1},
 issn = {2397-7132},
 journal = {Npj 2D Materials and Applications},
 doi = {10.1038/s41699-020-0155-x}
}

@article{Luria.2020,
 author = {Luria, Omer and Mohapatra, Pranab Kishore and Patsha, Avinash and Kribus, Abraham and Ismach, Ariel},
 year = {2020},
 title = {Large-Scale characterization of Two-Dimensional Monolayer \ce{MoS2} Island Domains Using Spectroscopic Ellipsometry and Reflectometry},
 pages = {146418},
 volume = {524},
 issn = {0169-4332},
 journal = {Applied Surface Science},
 doi = {10.1016/j.apsusc.2020.146418}
}

@article{Isic.2011,
 author = {Isic, Goran and Jakovljevic, Milka and Filipovic, Marko and Jovanovic, Djordje M. and Vasic, Borislav and Lazovic, Sasa and Puac, Nevena and Petrovic, Zoran Lj. and Kostic, Radmila and Gajic, Rados and Humlicek, Josef and Losurdo, Maria and Bruno, Giovanni and Bergmair, Iris and Hingerl, Kurt},
 year = {2011},
 title = {Spectroscopic ellipsometry of few-layer graphene},
 pages = {051809},
 volume = {5},
 number = {1},
 issn = {1934-2608},
 journal = {Journal of Nanophotonics},
 doi = {10.1117/1.3598162}
}

@article{Chen.2021,
 author = {Chen, Chao and Chen, Xiuguo and Wang, Cai and Sheng, Sheng and Song, Lixuan and Gu, Honggang and Liu, Shiyuan},
 year = {2021},
 title = {Imaging Mueller matrix ellipsometry with sub-micron resolution based on back focal plane scanning},
 pages = {32712--32727},
 volume = {29},
 number = {20},
 issn = {1094-4087},
 journal = {Optics Express},
 doi = {10.1364/OE.439941}
}

@article{Funke.2016,
 author = {Funke, S. and Miller, B. and Parzinger, E. and Thiesen, P. and Holleitner, A. W. and Wurstbauer, U.},
 year = {2016},
 title = {Imaging spectroscopic ellipsometry of \ce{MoS2}},
 pages = {385301},
 volume = {28},
 number = {38},
 issn = {0953-8984},
 journal = {Journal of Physics: Condensed Matter},
 doi = {10.1088/0953-8984/28/38/385301}
}

@article{Wurstbauer.2010,
 author = {Wurstbauer, Ulrich and R{\"o}ling, Christian and Wurstbauer, Ursula and Wegscheider, Werner and Vaupel, Matthias and Thiesen, Peter H. and Weiss, Dieter},
 year = {2010},
 title = {Imaging ellipsometry of graphene},
 pages = {231901},
 volume = {97},
 number = {23},
 issn = {0003-6951},
 journal = {Applied Physics Letters},
 doi = {10.1063/1.3524226}
}

@article{Ivanova.2024,
 author = {Ivanova, Tatyana V. and Andres-Penares, Daniel and Wang, Yiping and Yan, Jiaqiang and Forbes, Daniel and Ozdemir, Servet and Burch, Kenneth S. and Gerardot, Brian D. and Brotons-Gisbert, Mauro},
 year = {2024},
 title = {Optical contrast analysis of {$\alpha$}-\ce{RuCl3} nanoflakes on oxidized silicon wafers},
 pages = {071114},
 volume = {12},
 number = {7},
 issn = {2166-532X},
 journal = {APL Materials},
 doi = {10.1063/5.0212132}
}

@article{Puebla.2020,

 author = {Puebla, Sergio and Mariscal-Jim{\'e}nez, Antonio and Gal{\'a}n, Rosal{\'i}a S. and Munuera, Carmen and Castellanos-Gomez, Andres},
 year = {2020},
 title = {Optical-Based Thickness Measurement of \ce{MoO3} Nanosheets},
 volume = {10},
 number = {7},
 issn = {2079-4991},
 journal = {Nanomaterials},
 doi = {10.3390/nano10071272}
}

\end{document}